\pdfoutput=1 
\documentclass[aps,prl,10pt,twocolumn,superscriptaddress,longbibliography,nofootinbib,floatfix]{revtex4-2}
\usepackage{graphicx}
\usepackage{amsmath,amssymb}
\usepackage{xcolor}
\usepackage{hyperref}
\hypersetup{hidelinks}
\usepackage{xurl}
\usepackage[capitalize]{cleveref}

\DeclareFontFamily{U}{eur}{}
\DeclareFontShape{U}{eur}{m}{n}{<-6>eurm5<6-8>eurm7<8->eurm10}{}
\DeclareFontShape{U}{eur}{b}{n}{<-6>eurb5<6-8>eurb7<8->eurb10}{}
\DeclareFontShape{U}{eur}{bx}{n}{<->ssub*eur/b/n}{}
\newcommand{\eulername}[1]{\NoCaseChange{{\usefont{U}{eur}{m}{n}#1}}}
\newcommand{\qlaybench}{\texorpdfstring{\eulername{QlayBench}}{QlayBench}}
\newcommand{\qlaybot}{\texorpdfstring{\eulername{QlayBot}}{QlayBot}}

\newcounter{paperappendix}
\renewcommand{\thepaperappendix}{\Alph{paperappendix}}
\newcommand{\appendixparagraph}[1]{%
  \par\refstepcounter{paperappendix}%
  \textit{Appendix~\thepaperappendix. #1.}\enspace
}

\begin{document}

\title{Domain advantage in two-dimensional quantum-device design with multi-modal agentic harness}
\author{Shenhao Miao$^{*,\dagger}$}
\affiliation{GaugeForge PTE. LTD., 049422, Singapore}
\affiliation{Department of Condensed Matter Physics, Weizmann Institute of Science, Rehovot 7610001, Israel}
\author{Yizhou Li$^{*}$}
\affiliation{GaugeForge PTE. LTD., 049422, Singapore}
\author{Liqiao Xia}
\affiliation{Department of Physics, Massachusetts Institute of Technology, Cambridge, MA 02139, USA}
\author{Ruihao Liu}
\affiliation{Department of Physics, Massachusetts Institute of Technology, Cambridge, MA 02139, USA}
\author{Dingke Wang}
\affiliation{UHasselt, Institute for Materials Research (IUMAT), Hasselt 3500, Belgium}
\affiliation{imec, IUMAT, Diepenbeek 3590, Belgium}
\affiliation{EnergyVille, IUMAT, Genk 3600, Belgium}
\author{Pablo Jarillo-Herrero}
\affiliation{Department of Physics, Massachusetts Institute of Technology, Cambridge, MA 02139, USA}
\author{Eli Zeldov}
\affiliation{Department of Condensed Matter Physics, Weizmann Institute of Science, Rehovot 7610001, Israel}
\author{Jiaqi Cai$^{\dagger}$}
\affiliation{Department of Physics, Massachusetts Institute of Technology, Cambridge, MA 02139, USA}

\date{November 21, 2026}

\begin{abstract}
High-quality two-dimensional quantum devices are commonly fabricated from mechanically exfoliated crystals, whose irregular shapes and placement require a lithographic layout tailored to each specimen. Drawing these layouts remains largely manual. We introduce \qlaybench{}, a benchmark that tests whether language-model agents can turn microscope images and experimental requirements into device layouts that fit the specimen. Agents with conventional tools often construct devices when material outlines are supplied, but struggle to identify those outlines from images. Modelled on the human workflow, \qlaybot{}, a multimodal agentic harness, connects image interpretation to the tools used to align images, draw device features and inspect the resulting layout. We evaluate each layout with checks drawn from practical design experience, asking whether it fits the specimen and connects its intended components. With greater computational effort, an agent using \qlaybot{} passes 81.5\% of the benchmark tasks, compared with 24.1\% when it uses conventional tools. This domain advantage highlights the importance of connecting scientific interpretation of a specimen to geometric design and provides a route towards reducing manual effort in two-dimensional quantum-device fabrication.
\end{abstract}

\maketitle

{\let\thefootnote\relax\footnotetext{$^{*}$~These authors contributed equally to this work.}}
{\let\thefootnote\relax\footnotetext{$^{\dagger}$~Corresponding authors, shenhao.miao@gauge-forge.com (S.M.) and jiaqi.cai@gauge-forge.com (J.C.)}}

\begin{figure*}[t]
\centering
\includegraphics[width=\textwidth]{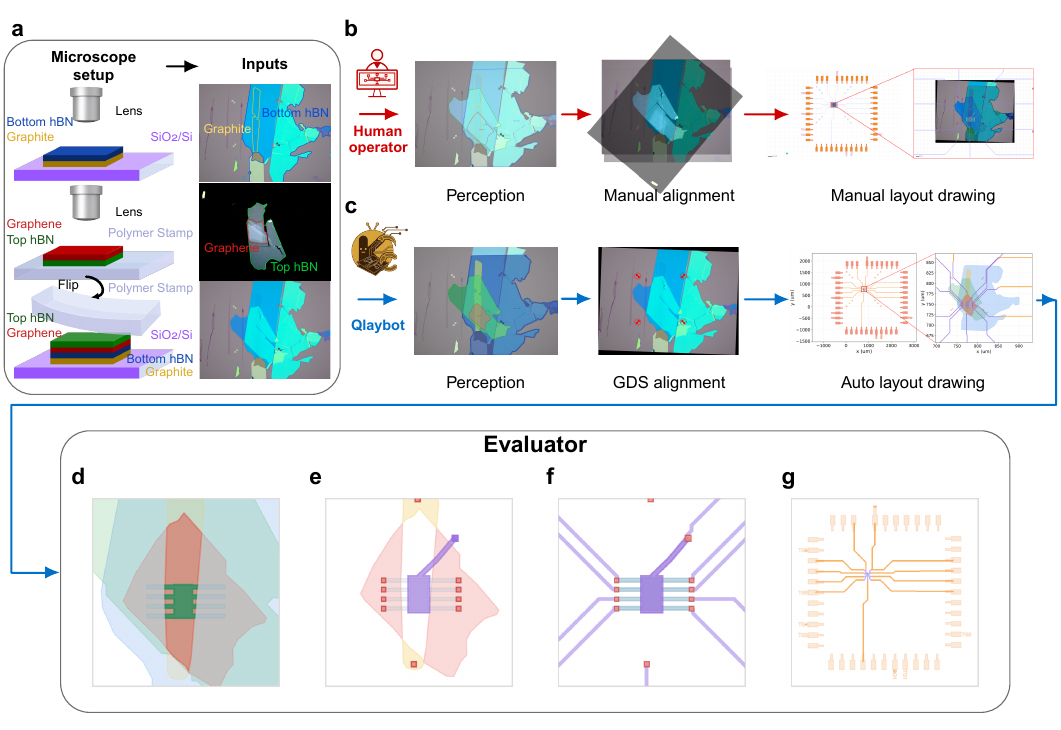}
\caption{\textbf{From the human design workflow to an agent harness and evaluator.}
(a)~Assembly and microscope inputs. Bottom hBN (blue outline) on graphite (yellow outline) is imaged on a SiO$_2$/Si chip. Top hBN (green outline) and graphene (red outline) are imaged on a polymer stamp before flipping and transfer onto the backgate. A third image records the assembled heterostructure.
(b)~A human operator identifies the material layers, manually aligns the images to the layout coordinates and draws the device in EDA software.
(c)~With \qlaybot{}, the agent emulates this sequence with computer-assisted perception, marker-based alignment and geometric design tools, including an autorouter.
(d--g)~Evaluation checkpoints abstracted from practical design experience. The evaluator checks the active region within the material overlap (d), contacts and gates on the intended layers (e), local lead routing (f) and connections to the bonding pads (g).}
\label{fig:task}
\end{figure*}

\section{Introduction}\label{sec:intro}
Mechanically exfoliated two-dimensional (2D) crystals provide some of the cleanest platforms for studying electronic phases in reduced dimensions \cite{Novoselov2004Graphene,Novoselov2005Dirac,Zhang2005QHE,Cao2018MagicAngle}. Unlike conventional epitaxial heterostructures, atomically thin crystals can be assembled without strict lattice matching, allowing materials with distinct electronic \cite{Novoselov2004Graphene,Dean2010hBN,Cao2018MagicAngle,Cao2018CorrelatedInsulator,Cai2023FQAH,Zeng2023FCI,Park2023FQAH,Xu2023FQAH,Lu2024FQAH}, magnetic \cite{Huang2017CrI3,Gong2017CGT,Telford2020CrSBr,Auerbach2025Isospin} and optical properties \cite{Cao2012Valley,Xiao2012Valleytronics} to be tailored in a single structure \cite{Geim2013VdW,Novoselov2016VdW}. This flexibility has made van der Waals heterostructures an important platform for quantum states and device concepts that are difficult to realize in bulk materials \cite{Liu2016VdWDevices,Balents2020Moire,Andrei2021Moire,MakShan2022Moire}. However, exfoliated crystals have irregular boundaries and stochastic positions, and their relative placement changes during assembly. Because the final geometry is known only after exfoliation and stacking, the lithographic mask cannot be finalized in advance \cite{Novoselov2004Graphene,Wang2013EdgeContact}. Each layout is therefore usually drawn manually in electronic design automation (EDA) software, and the effort grows with the number of material interfaces, contacts and fabrication constraints.

Automation has advanced rapidly in van der Waals device fabrication. Computer-vision methods have been developed to identify and segment exfoliated crystals in optical microscope images \cite{Masubuchi2020Segmentation}, while robotic platforms can search for, transfer and assemble large numbers of crystals into heterostructures \cite{Masubuchi2018Robotic,Mannix2022Robotic,Li2026RoboticAssembly,Shi2026Qumus}. These developments automate the search for suitable crystal flakes and the stacking of heterostructures. However, the design of a measurement device around the resulting structure is still largely manual. The designer must connect microscope images with experimental requirements. Microscope images supply the available material geometry, while experimental specifications determine where the active region, contacts and gates should lie. A useful design must satisfy both these physical requirements and the geometric constraints of fabrication.

Language-model agents provide a promising interface between experimental specifications and professional design software \cite{Yao2023ReAct,Schick2023Toolformer,Shinn2023Reflexion}. Agents that process images and text can operate graphical interfaces and call external software tools, but computer-use and scientific benchmarks expose difficulties with precise and extended workflows \cite{Xie2024OSWorld,Ai2026ProSoftArena,Chen2025ScienceAgentBench,Siegel2024COREBench}. In atomic force microscopy, for example, performance on materials-science questions does not reliably predict success in experimental automation \cite{Mandal2025AFM}. Recent work in computer-aided design connects language and vision models to tools that construct shapes, display them for inspection and check their geometry \cite{Doris2026CADBench,Barkley2026CADSmith,Gong2026ToolCAD}. These studies primarily address the generation or reconstruction of three-dimensional mechanical parts from textual descriptions or target geometries. In contrast, a 2D quantum device must be designed around a unique experimental specimen, and interpreting that specimen requires scientific understanding. Several layouts can satisfy the requested function. An evaluator must therefore test how the design fits the material geometry and connects the device to its electrodes.

We formulate this problem as \qlaybench{}, a benchmark that tests interpretation of microscope images, device construction and their end-to-end combination. Inspired by the human workflow, we embed checkpoints in the design workflow and test the agent at each one, which exposes a practical bottleneck. Agents that can construct devices from supplied material contours often fail when they must infer those contours from images. To remove this bottleneck, we introduce \qlaybot{}, a multimodal agentic harness that gives agents computer-vision tools, geometric operations and visual feedback within a workflow modelled on human design practice. We use the term \emph{domain advantage} for the higher task success achieved with \qlaybot{} relative to the conventional model--harness combinations evaluated here. Together, \qlaybot{} and \qlaybench{} suggest that multimodal harness engineering, rather than model capability alone, decides whether an agent can design around a real specimen, and they set a benchmark against which future agents for two-dimensional device design can be measured.

\section{Setup}\label{sec:setup}
\Cref{fig:task} summarizes the task setting. We consider the two-step fabrication procedure used in much of two-dimensional material research \cite{Wang2013EdgeContact,Cao2018MagicAngle}, shown in \cref{fig:task}(a). In the first step, a heterostructure of hexagonal boron nitride (hBN), a two-dimensional dielectric, on top of a graphite electrode is prepared on a SiO$_2$/Si wafer chip. This stack provides the backgate and its dielectric for electrostatic gating \cite{Dean2010hBN,Zibrov2017GraphiteGate,Overweg2018QPC}. The chip also carries the alignment markers that anchor the electronic design and a fixed bonding-pad template. A microscope image of this stage forms the first input. The layered materials and a target area can be identified by eye. In the second step, a polymer stamp carries the top hBN together with the target material, such as graphene. An image of this structure before transfer forms the second input. The stamp is flipped and the top structure is laminated onto the backgate, completing the heterostructure and producing the third image. These three images and the bonding-pad template of the chip are the geometric inputs to the end-to-end design task. Image scales and material identities are supplied, but aligned material outlines are not.

In the conventional workflow of \cref{fig:task}(b), a human operator turns these inputs into a layout in three stages. The first stage is perception. The operator inspects the microscope images and identifies each layer, from the buried graphite to the target material on the stamp, sometimes using a contrast tool to enhance features. The operator then manually registers the alignment markers in the images to the design file. In the final stage, the operator draws the polygons of each layout layer by hand in EDA software.

The \qlaybot{} workflow of \cref{fig:task}(c) closely mimics this human procedure. The agent uses its ability to interpret images \cite{OpenAI2023GPT4V,Google2023Gemini} and computer-vision tools to identify material layers and registers the images to layout coordinates through the alignment markers. It constructs the device with geometric tools and an autorouter, which generates lead paths from device contacts to the fixed bonding pads, in KLayout, an open-source EDA tool for viewing and editing layouts in electronics and photonics \cite{Koefferlein2024KLayout}. Both workflows produce a layered layout in Graphic Data System II (GDSII), a standard format for lithographic masks. An automated evaluator scores the submitted layout against the geometric requirements of the task.

We also draw on the human workflow to define evaluation checkpoints. In practical design, identifying the material layers is only the first step. An operator must then place the active region within their overlap, connect the intended layers and carry those connections to the bonding pads. A design can be correct at one length scale and still fail at the next. We formalize these checks in an automated evaluator at four scales (\cref{fig:task}(d--g)). At the flake scale, it measures how much of the active region lies inside the overlap of the required layers. At the device scale, it assesses contact placement and gate coverage. At the routing scale, it checks crossings and connections between leads. At the chip scale, it verifies connections to the bonding-pad frame. The harness thus follows the operator's sequence of actions, while the evaluator tests the geometric requirements that motivate those actions.

\section{\qlaybench{}}\label{sec:qlaybench}
\begin{figure*}[t]
\centering
\includegraphics[width=\textwidth]{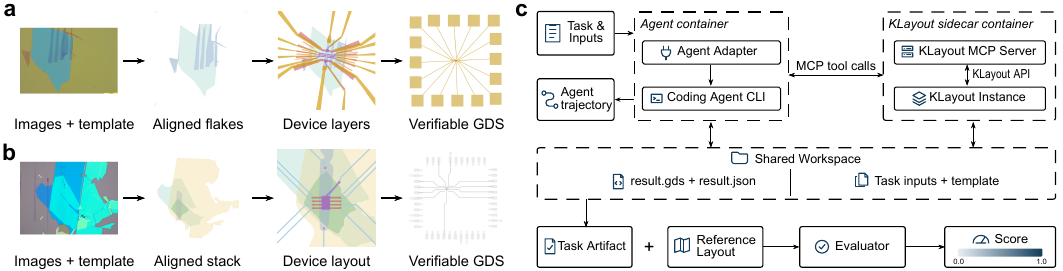}
\caption{\textbf{Benchmark tasks and execution.}
(a,b)~Two representative instances, each progressing from microscope images and a bonding-pad template to aligned material contours, device layers and a complete GDSII layout. The three task families test recovery of the aligned contours, design from supplied contours and the complete workflow.
(c)~The agent uses design software to save a layout, which is checked against independently annotated specimen boundaries outside its workspace. CLI denotes command-line interface; API, application programming interface; and MCP, Model Context Protocol. GDS denotes the GDSII layout file format.}
\label{fig:tasks}
\end{figure*}

\qlaybench{} uses these checkpoints to test where agents succeed or fail in the design workflow. The benchmark comprises 54 tasks drawn from 18 fabricated heterostructures (Appendix~\ref{app:benchmark}), each specifying an eight-terminal Hall bar with gate and backgate connections. Each heterostructure defines three task families. Flake detection asks for the constituent material contours in layout coordinates. Device design supplies aligned material geometry and asks for a complete layout. End-to-end design requires the full conversion from microscope images and experimental specifications to a layout. \Cref{fig:tasks}(a,b) illustrates the intermediate geometries and final outputs for two representative specimens.

The task families place checkpoints between perception and geometric design without requiring a unique final drawing. Flake detection is scored by weighted overlap with reference contours drawn by experienced operators. Device and end-to-end design are scored on the geometric and connectivity requirements summarized in \cref{fig:task}(d--g), so different layouts can pass. A task passes when its score meets a family-specific threshold (Appendix~\ref{app:evaluation}). These thresholds differ across families, and a contour error that fails the segmentation criterion can still leave the chosen channel and contacts within their intended regions. The family pass rates therefore measure different requirements along the workflow.

\Cref{fig:tasks}(c) shows the execution environment. Each agent receives the task inputs and saves its completed layout. With \qlaybot{}, it uses image-processing and layout tools through a shared design workspace. The conventional configurations use their coding interfaces without this tool layer. Evaluation is performed outside the agent workspace using reference geometry unavailable to the agent during the trial. We compare seven model--harness combinations, comprising six conventional configurations and GPT-5.6-sol using \qlaybot{}. For each combination, we report one scored submission per task.

\begin{figure*}[t]
\centering
\includegraphics[width=\textwidth]{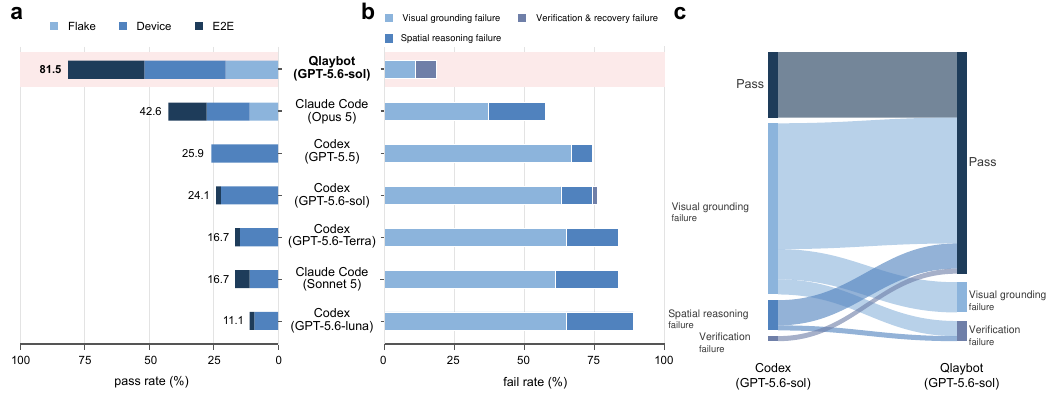}
\caption{\textbf{Task success and failure modes.}
(a)~Pass rates for seven model--harness combinations. Stacked segments give the contributions of flake detection (Flake), device design (Device) and end-to-end design (E2E) to the total, each calculated as the number of passed tasks in that family divided by all 54 benchmark tasks. The total bar length is therefore the overall pass rate, rather than the sum of the three within-family pass rates.
(b)~Failure rates under this denominator, grouped by problems with identifying and aligning materials (visual grounding), constructing the device (spatial reasoning), or detecting and correcting errors (verification and recovery). The bar totals in (a) and (b) sum to 100\% for each combination.
(c)~Outcomes for corresponding tasks with GPT-5.6-sol under the conventional Codex interface and \qlaybot{}. Of 31 failures that become passes, 25 were labelled as visual grounding failures.}
\label{fig:bench}
\end{figure*}

\Cref{fig:bench} compares task success, failure modes and the change in outcomes when GPT-5.6-sol is equipped with the \qlaybot{} harness. The conventional combinations pass between 6 and 23 of the 54 tasks (11.1--42.6\%), with Claude Opus 5 under Claude Code achieving the highest total (\cref{fig:bench}(a)). Their performance differs sharply across task families. GPT-5.5 designs 14 of 18 devices when aligned contours are supplied, but passes no flake-detection or end-to-end tasks. GPT-5.6-sol passes 12 of 18 device-design tasks, no flake-detection tasks and only one end-to-end task. Five of the six conventional combinations pass no flake-detection tasks. These results indicate a bottleneck at the conversion of microscope images into material geometry that downstream design operations can use.

Using \qlaybot{}, GPT-5.6-sol passes 44 of 54 tasks (81.5\%), compared with 13 of 54 (24.1\%) under the conventional Codex interface (\cref{fig:bench}(a)). The family totals change from 0 to 11 of 18 for flake detection, from 12 to 17 for device design and from 1 to 16 for end-to-end design. The gain is largest in the complete image-to-layout workflow, where each geometric operation depends on the output of perception and registration. This model--harness combination also exceeds the 42.6\% overall pass rate of the strongest conventional combination.

The failure analysis supports this interpretation (\cref{fig:bench}(b)). We assign each failed submission to visual grounding, spatial reasoning or verification and recovery using the recorded actions and evaluator output (Appendix~\ref{app:failures}). Visual grounding includes material identification, contour selection and image alignment. Spatial reasoning concerns device placement and connections, while verification and recovery concerns detecting and correcting errors. Visual grounding accounts for 193 of the 250 conventional failures and is the largest failure class for every conventional combination. Supplying aligned contours bypasses these operations, allowing us to distinguish difficulty with the experimental images from difficulty constructing the device itself.

The ten residual failures with \qlaybot{} comprise six visual grounding cases and four verification and recovery cases. In one of the latter, the agent submitted contours despite a check indicating that the material contours had not been aligned correctly. In the other three, internal checks reported a successful layout that the evaluator rejected. Tools can therefore expose a problem without ensuring that the agent corrects it, and a geometrically consistent drawing can still be misplaced relative to the material.

\Cref{fig:bench}(c) follows the outcomes of corresponding tasks under conventional Codex and \qlaybot{}. Across the 54 tasks, 31 failures become passes, all 13 previous passes are retained and ten tasks remain failures. Of the 34 conventional failures labelled as visual grounding, 25 become passes with \qlaybot{}. The transition therefore occurs predominantly among tasks that previously failed to recover usable material geometry. Eight end-to-end tasks are passed only with \qlaybot{} among the evaluated combinations.

This comparison measures the performance of the implemented systems under their recorded execution protocols. The record for GPT-5.6-sol using \qlaybot{} combines recovery runs, whereas its conventional record comes from a single run. Mean execution time and recorded cost for the runs producing the reported submissions are also higher with \qlaybot{}, by about factors of five and seven, respectively (Appendix~\ref{app:cost}). All reported submissions were evaluated against the expert reference annotations (Appendix~\ref{app:configurations}). The comparison establishes the gain for the complete harness under the evaluated protocol; isolating individual tool contributions or comparing at equal computational expenditure requires separate experiments.

\section{\qlaybot{}}\label{sec:qlaybot}
\begin{figure*}[t]
\centering
\includegraphics[width=\textwidth]{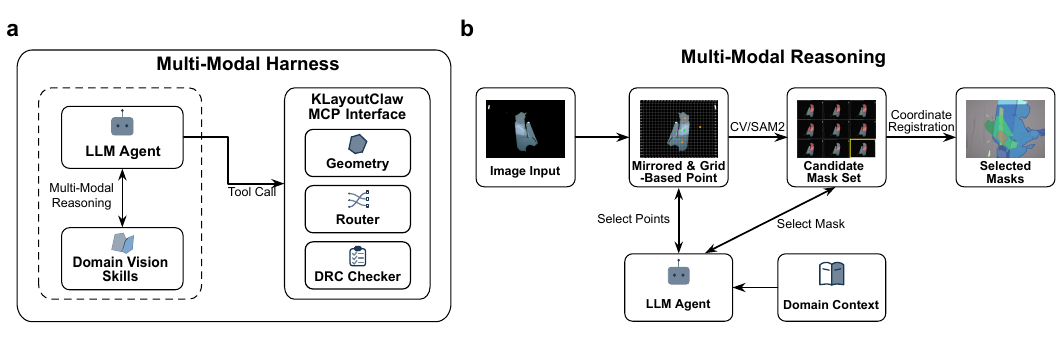}
\caption{\textbf{The \qlaybot{} multimodal harness.}
(a)~A large language model (LLM) agent uses written guidance to interpret specimen images and uses KLayoutClaw, a set of tools for editing the layout, to construct geometry, route leads and check the resulting geometry.
The domain vision skills provide written material and assembly guidance. Through the Model Context Protocol (MCP) interface, geometric tool calls edit device features, the router generates leads, and the design-rule checker (DRC Checker) checks layout geometry.
(b)~The perception workflow expands the multimodal reasoning component in (a). A micrograph is mirrored when required by the transfer geometry and overlaid with a grid. The agent uses material context to select points that guide the segmentation tool, inspects the proposed material outlines and selects material contours. Image alignment places these contours in the layout frame. CV denotes computer vision; SAM 2 denotes Segment Anything Model 2. The CV/SAM2 box is the segmentation tool that generates candidate masks from the selected points. Coordinate registration maps the selected contours into the layout frame. The model selects and interprets candidates, while the vision tools extract their boundaries and align them with the layout.}
\label{fig:qlaybot}
\end{figure*}

We now examine how \qlaybot{} connects image interpretation to device construction. \Cref{fig:qlaybot}(a) shows its two components. Written guidance about the materials and assembly process leads the agent through material identification and image alignment. KLayoutClaw, a Model Context Protocol (MCP) server that exposes KLayout's editing functions to the agent \cite{Anthropic2024MCP}, provides tools to draw device features, connect them to the bonding pads and check the resulting geometry. The agent selects the material regions and design operations, while the tools extract boundaries, align images and edit the layout. This division of work follows the human sequence of perception, alignment and drawing introduced in \cref{fig:task}(b,c).

The geometric tools act on the evolving layered layout. Layer-aware operations let the agent construct device features relative to the material contours, and the autorouter connects device contacts to the fixed bonding-pad template. Structural design-rule checks examine the resulting geometry. Rendered views let the agent inspect where its edits lie relative to the material and to features drawn earlier. The agent can then revise the design through further tool calls. These visual and geometric checks support the design process, while the external evaluator independently tests the submitted layout against reference geometry.

\Cref{fig:qlaybot}(b) details how image interpretation supplies the geometry on which these operations depend. The first step accounts for the assembly sequence. A micrograph taken before the stack is flipped is mirrored when required to match the final device orientation. The image is then overlaid with a reference grid, allowing the agent to mark points inside the materials it identifies. The agent thus chooses regions, not vertices. The segmentation tool (CV/SAM2 in \cref{fig:qlaybot}(b)) uses these points to generate candidate segmentation masks, from which material outlines are extracted. The agent inspects these candidates using the image and its understanding of the material stack, then selects the outlines corresponding to the intended layers. This selection retains the scientific interpretation required by the task. Visual contrast alone singles out a region, while only knowledge of the material stack identifies it as the target material and places its boundary on the correct layer. The model therefore supplies material identity and context, while the segmentation tool determines the candidate boundaries in image coordinates.

Coordinate registration (\cref{fig:qlaybot}(b)) places the selected outlines in the coordinate frame used for fabrication. The source images are first aligned with the assembled stack using recognizable image features and flake outlines. Chip markers then locate these contours in the layout, where they provide the material geometry for device construction (Appendix~\ref{app:implementation}). This step connects the two panels of \cref{fig:qlaybot}. The overlap between registered layers determines the available active region, and the contours constrain contact placement and routing to the bonding pads. By carrying the selected material geometry into the design environment, the harness makes the agent's interpretation of the specimen available to subsequent geometric operations and visual checks.

\section{Discussion}\label{sec:discussion}
The principal result is that an agent designs far better when it selects material contours than when it must produce them. Multimodal language models recognize materials in a micrograph yet localize their boundaries poorly \cite{Rahmanzadehgervi2024Blind,Fu2024BLINK,Tong2024EyesWideShut}. Their ability to generate an output is also distinct from their ability to judge one \cite{West2024Paradox}, and selection among given candidates is the more reliable of the two \cite{Saunders2022SelfCritique}. \qlaybot{} therefore lets the agent mark regions and choose among tool-proposed outlines \cite{Kirillov2023SAM,Yang2023SoM,Lei2025Scaffold}, which offloads the precise spatial work to the tool \cite{Risko2016Offloading} and leaves the model with the scientific interpretation. Our checkpoint design is an ablation of the workflow. Supplying aligned contours removes the perception step, and the gap between the device-design and image-based task families measures the burden that step carries. Contacts, gates and leads may be internally consistent yet unusable if the inferred material boundaries or coordinate transform are wrong. \qlaybot{} resolves this dependence by carrying the material contours the agent selects from perception, through registration, into the design environment. Its domain advantage on \qlaybench{} supports treating this connection as an essential part of the agent's capability.

A second result is that a passing score certifies only the requirement it measures, and correctness of the whole layout requires an independent reference to the specimen. The flake-detection score rewards agreement with the specimen contours, whereas the device score rewards placement and connectivity of the chosen geometry, so an agent can pass an end-to-end task while failing flake detection for the specimen. The residual verification failures with \qlaybot{} show the opposite case. Because an agent's own checks use the geometry it inferred, only an evaluator holding an independent reference to the specimen can reject a layout built on the wrong contours.

One design rule suggested by the benchmark result is to let the agent select the geometry of the specimen rather than draw it, and to carry that selection into the coordinates used to construct and check the device. This principle applies more broadly to scientific design tasks in which each specimen has a different geometry: the model uses scientific context to identify relevant regions and select among tool-generated candidates, while specialized tools determine precise boundaries and coordinate transformations. Carrying the selected geometry into subsequent operations preserves the connection between interpretation, construction and verification. This division of labour offers a design strategy for other image-based scientific workflows in which semantic interpretation must guide precise spatial operations. \qlaybot{} implements this division of labour by emulating the human design workflow, and \qlaybench{} turns the checkpoints of that workflow into an external test of the resulting layouts. The approach complements robotic crystal assembly \cite{Masubuchi2018Robotic,Mannix2022Robotic,Li2026RoboticAssembly,Shi2026Qumus} by addressing the specimen-specific lithographic design stage that follows stacking. Together, the harness and evaluator provide a route towards reducing manual design effort while checking the geometric conditions that turn an assembled heterostructure into a measurement device.

\emph{Data and code availability.} Three representative data entries are provided for verification. Additional data supporting the findings of this study are available from the corresponding authors upon reasonable request. The accompanying Supplemental Material gives the evaluator specification, implementation details and specimen list. KLayoutClaw tool implementations and skill texts are available at \url{https://github.com/caidish/KlayoutClaw}.


\emph{Acknowledgments.} E.Z. acknowledges the support of the Knell Family Institute for Artificial Intelligence. Work in the P.J.-H. lab was supported by the MIT Generative AI Impact Consortium, the MIT/Microsystems Technology Laboratories Samsung Semiconductor Research Fund, the MIT-Portugal Program, the Gordon and Betty Moore Foundation's EPiQS Initiative through Grant GBMF9463 to P.J.-H., the Heising-Simons Foundation, and the Ramon Areces Foundation. All work by J.C. was conducted at MIT, and J.C. acknowledges support from the MIT Pappalardo Fellowship in Physics.

\bibliography{main}

\section*{End Matter}
\appendixparagraph{Benchmark construction}\label{app:benchmark} The benchmark contains 18 heterostructures supplied by researchers at the Massachusetts Institute of Technology. Each specimen contributes three tasks, so the 54 tasks represent 18 distinct material geometries. All contributed specimens meeting the task requirements were included, without selection based on agent performance. The device tasks request an eight-terminal Hall bar with gate and backgate connections. Thirteen specimens have graphite backgates and five have gold backgates. Figure~\ref{fig:task} illustrates the graphite arrangement. Supplemental Material, Sec.~\ref{sec:supp-qlaybench} lists the specimens and task definitions.

The specimen providers and authors jointly annotated the material boundaries using the optical and polarized microscope images, with knowledge of the material identities and assembly history.

\appendixparagraph{Evaluation}\label{app:evaluation} Several layouts can satisfy the requirements of a particular specimen. To express this objective, let $I$ denote the microscope images, $M$ the material geometry, $s$ the requested device specification and $B$ the bonding-pad template. The agent infers material geometry $\widehat M$ from the images and constructs a layout $L$. Let $C_j$ equal one when requirement $j$ is satisfied and zero otherwise, and let $\mathcal H$ list the mandatory requirements. The set of acceptable layouts is
\begin{equation}
\mathcal F(M,s,B)=\{L\mid C_j(L,M,s,B)=1\ \text{for all }j\in\mathcal H\},
\label{eq:feasible}
\end{equation}
This expression states the physical objective. The evaluator combines geometric checks into a score measuring agreement with the design requirements.

For flake detection, each inferred outline is compared with an expert outline by dividing their common area by the area covered by either outline. This intersection-over-union measures agreement from zero to one. Each material receives zero credit below 0.4, half credit from 0.4 to below 0.7, and full credit from 0.7. Graphene and the backgate each carry twice the weight of either hBN layer. Only materials present in the reference contribute to the average.

For device design, ten checks measure channel placement, contact placement, gate coverage and the geometry of the connections. Weights reflect the checks' influence on device function, and thresholds reflect typical human performance. The passing thresholds are 0.80 for flake detection, 0.95 for design from supplied outlines and 0.80 for end-to-end design, with equality accepted. Supplemental Material, Sec.~\ref{sec:supp-qlaybench} defines the evaluation terms and passing thresholds.

\appendixparagraph{Agent configurations}\label{app:configurations} The conventional configurations use four GPT models through the Codex command-line interface and two Claude models through Claude Code, as named in Fig.~\ref{fig:bench}. These are interfaces through which a model reads files, runs programs and submits its work. GPT-5.6-sol also uses \qlaybot{}.

All reported submissions were evaluated against the expert reference annotations. The evaluator runs outside the agent workspace and uses reference annotations unavailable to the agent during execution.

\appendixparagraph{Execution time and cost}\label{app:cost} For the runs producing the reported submissions, conventional GPT-5.6-sol averages 12.7 minutes and \$1.71 per task. With \qlaybot{} the corresponding means are 63.3 minutes and \$12.11.

\appendixparagraph{\qlaybot{} implementation}\label{app:implementation} The agent follows written instructions about material interpretation, image alignment, device construction and repair. KLayoutClaw supplies the tools for editing and checking layouts in KLayout \cite{Koefferlein2024KLayout}. A KLayout process running alongside the agent receives its requests through the Model Context Protocol, a standard interface for calling external tools \cite{Anthropic2024MCP}. This arrangement lets the agent modify a live layout and inspect the result.

Image-processing tools propose material boundaries and the model chooses among them. For graphene and graphite, the Segment Anything Model~2, an image-segmentation model guided by selected points, can propose several candidate regions \cite{Ravi2024SAM2}. Conventional image-processing routines provide additional candidates and identify hBN boundaries.

The alignment tools first place the source images in the assembled-stack view, then place that view in chip coordinates using alignment markers. Layout tools compute the material overlap and the regions belonging to only one conductive layer. The model chooses device features within these regions and supplies the contacts, pads and constraints to the routing tool. It can also specify connections directly. Checks return measurements and locations of problems, but the model decides whether to revise the design or submit it. Supplemental Material, Sec.~\ref{sec:supp-qlaybot} details the image-registration and routing procedures.

\appendixparagraph{Failure analysis and specimen examples}\label{app:failures} We assign one principal failure category after reading the agent's recorded actions and evaluator output. Visual grounding covers identifying materials, choosing their outlines and aligning them with the layout. Spatial reasoning covers placement of device features and the arrangement of connections. Verification and recovery covers missed errors, unresolved warnings and disagreement between the agent's checks and external evaluation.

Two examples in Supplemental Material, Sec.~\ref{sec:supp-designs} compare the material outlines and device geometry from individual completed designs. The specimen identifiers QH06 and ml14 are archive labels. In QH06, an agent using Claude Opus 5 overestimates the graphene boundary, yet the channel and eight contact centres occupy their intended reference regions. In ml14, an agent using Claude Sonnet 5 accepts contact positions using its own inferred boundaries, although all eight contact patches lie outside both reference conductive materials.

Let $V_{\rm internal}$ denote checks using the agent's inferred geometry $\widehat M$, and $V_{\rm reference}$ checks using independently annotated geometry $M_{\rm ref}$. For a layout $L$, the checks take different material geometry as their input
\begin{equation}
V_{\rm internal}(L,\widehat M)\quad\text{and}\quad
V_{\rm reference}(L,M_{\rm ref}),
\label{eq:verification}
\end{equation}
Internal checking can share the mistaken geometry used to construct the device.

%
\clearpage
\onecolumngrid
\raggedbottom 

\setcounter{secnumdepth}{2}
\setcounter{section}{0}
\setcounter{figure}{0}
\setcounter{table}{0}
\setcounter{equation}{0}
\renewcommand{\thesection}{S\arabic{section}}
\renewcommand{\thesubsection}{\thesection.\arabic{subsection}}
\makeatletter
\renewcommand{\p@subsection}{}
\makeatother
\renewcommand{\thefigure}{S\arabic{figure}}
\renewcommand{\thetable}{S\arabic{table}}
\renewcommand{\theequation}{S\arabic{equation}}
\renewcommand{\theHsection}{S\arabic{section}}
\renewcommand{\theHsubsection}{S\arabic{section}.\arabic{subsection}}
\renewcommand{\theHfigure}{S\arabic{figure}}
\renewcommand{\theHtable}{S\arabic{table}}
\renewcommand{\theHequation}{S\arabic{equation}}

\pdfbookmark[1]{Supplemental Material}{supplemental-material}
\begin{center}
{\large\textbf{Supplemental Material}}\\[6pt]
\textbf{Domain advantage in two-dimensional quantum-device design with multi-modal agentic harness}
\end{center}
\label{supplemental-material}

\section{\qlaybench{}: specimens, tasks, and evaluation}\label{sec:supp-qlaybench}

\subsection{Specimens and task definitions}\label{sec:supp-specimens}

Specimen identifiers and backgate materials are listed below.

\begin{center}
\begin{tabular}{ll}
\hline\hline
Backgate & Specimen identifiers \\
\hline
Graphite (13) & ml04, ml08, ml09, ml11, ml14, AH02, AH03, AH06, AH07, HM08, QH06, QH07, QH12 \\
Gold (5) & AH05, HM05, HM06, HM07, HM11 \\
\hline\hline
\end{tabular}
\end{center}

Task packages contain microscope images from the assembly stages: bottom
hBN on the backgate, top hBN and graphene on a transfer stamp, and the
assembled stack. They specify material identities, image scales,
transfer-reflection conventions, the chip's bonding-pad template, and an
eight-terminal Hall-bar design with topgate and backgate connections.

Each specimen contributes three tasks. Flake detection submits aligned
contours for the constituent layers. Device design constructs a GDSII
layout from supplied contours. End-to-end design infers the contours
from microscope images and generates the layout. The evaluator compares
each submission with expert reference geometry held outside the agent
workspace.

\subsection{Flake detection}\label{sec:supp-detection}

The evaluator associates four material roles with the submitted GDS
layers: top hBN (10/0), graphene (11/0), bottom hBN (12/0), and the
conductive backgate (13/0). For material \(l\), let \(P_l\) and \(R_l\)
denote the predicted and reference regions. Their
intersection-over-union is
\[
I_l=\frac{|P_l\cap R_l|}{|P_l\cup R_l|}.
\]
The evaluator assigns a tiered score \(t(I_l)\) to each
reference-present material:
\[
t(I_l)=\begin{cases}
0, & I_l<0.4,\\
0.5, & 0.4\le I_l<0.7,\\
1, & I_l\ge 0.7.
\end{cases}
\]
Graphene and the backgate have weight 2; the two hBN layers have weight
1. The flake-detection score is
\[
S_{\mathrm{det}}=\frac{\sum_{l\in\mathcal L}w_l\,t(I_l)}{\sum_{l\in\mathcal L}w_l},
\qquad (w_{\mathrm{top\ hBN}},w_{\mathrm{Gr}},w_{\mathrm{bottom\ hBN}},w_{\mathrm{BG}})=(1,2,1,2),
\]
where \(\mathcal L\) contains the materials present in the reference.
Detection passes at \(S_{\mathrm{det}}\ge 0.80\).

\subsection{Device geometry and scoring}\label{sec:supp-device}

The device evaluator reads the mesa, contact patches, topgate, lead
routes, and bonding pads from the component layers specified by the
submitted layout map. Let \(M\) denote the mesa, \(G\) the reference
graphene region, and \(B\) the reference backgate. The active channel
lies in \(G\cap B\). Graphene-only contacts occupy \(G\setminus B\), and
backgate contacts occupy \(B\setminus G\).

The evaluator estimates the central channel \(C(M)\) using two geometric
constructions to accommodate mesas with different channel aspect ratios
and connecting-arm shapes. A distance-transform candidate uses a
weighted principal-axis fit to identify the broad interior. A
skeleton-based candidate follows the central spine and reconstructs an
interior region. The evaluator selects the candidate with greater
compactness, \(4\pi A/P^2\), where \(A\) and \(P\) are its area and
perimeter. Device checks use \(C(M)\) to distinguish the channel from
connecting arms.

Contact patches are classified from their overlap with the exclusive
material regions and the topgate. Material eligibility requires the
patch center to lie in the region, or an overlap of at least
\(\max(0.1\ \mu\mathrm{m}^2,0.1 A_{\mathrm{patch}})\). Mesa adjacency
uses a \(0.5\ \mu\mathrm{m}\) tolerance. The ten scoring terms evaluate
the resulting geometry:

\begin{center}
\begin{tabular}{lrl}
\hline\hline
Term & Weight & Geometric measurement \\
\hline
\(s_1\) Channel placement & 0.15 & Fraction of \(C(M)\) inside \(G\cap B\) \\
\(s_2\) Contact material & 0.10 & Eligible contacts, normalized by expected arm count \\
\(s_3\) Gate geometry & 0.10 & Channel coverage and isolation from unintended routes \\
\(s_4\) Route crossings & 0.10 & Penalty for lead intersections outside accepted junctions \\
\(s_5\) Contact-to-pad routing & 0.10 & Maximum distinct contact--pad matching along connected routes \\
\(s_6\) Route endpoints & 0.05 & Fraction of endpoints attached to device features or pads \\
\(s_7\) Mesa adjacency & 0.10 & Eligible contacts adjoining the mesa \\
\(s_8\) Arm connections & 0.10 & Contacts attached through mesa arms to the material overlap \\
\(s_9\) Lead separation & 0.10 & Number of connected route components relative to terminals \\
\(s_{10}\) Material isolation & 0.10 & Penalty for lead overlap with the opposite conductive material \\
\hline\hline
\end{tabular}
\end{center}

For example, the channel-placement term is
\[
s_1=\frac{|C(M)\cap G\cap B|}{|C(M)|},
\]
and the contact-to-pad term uses the maximum number \(K\) of distinct
matched contact--pad pairs,
\[
s_5=\frac{K}{\max(N_{\mathrm{eligible}},E)},
\]
where \(N_{\mathrm{eligible}}\) counts eligible contacts and \(E\)
denotes the expected arm count, nominally eight. The gate term combines
channel coverage and an intersection-based isolation check with equal
weights. The route-crossing term assigns full credit to zero counted
crossings and decreasing credit as crossing count increases. The
remaining terms use geometric counts, distances, and overlap penalties
as indicated in the table.

The evaluator combines the ten terms and applies a connectivity penalty:
\[
\begin{aligned}
Q={}&0.15s_1+0.10(s_2+s_3+s_4+s_5)\\
&+0.05s_6+0.10(s_7+s_8+s_9+s_{10}),\\
S_{\mathrm{device}}={}&Q\times\begin{cases}0.5,&s_5<0.5,\\1,&s_5\ge 0.5.\end{cases}
\end{aligned}
\]
The supplied-contour device task passes at
\(S_{\mathrm{device}}\ge0.95\); the end-to-end task passes at
\(S_{\mathrm{device}}\ge0.80\).

\section{\qlaybot{}: image interpretation and layout construction}\label{sec:supp-qlaybot}

\subsection{Material selection}\label{sec:supp-selection}

\qlaybot{} connects a multimodal language-model agent to image-processing
tools and a running KLayout editor through KLayoutClaw, supplying task
images, physical scales, material identities, device requirements, and
written design guidance. The agent selects material contours, places
device features, invokes layout operations, and revises the design after
inspection.

Conventional image-processing routines propose boundaries from optical
contrast, color, and connected regions, including hBN boundaries. For
graphene and graphite, the agent can also guide Segment Anything Model 2
(SAM2) with positive and negative image points. SAM2 returns candidate
masks and contours, which the agent inspects against the source image
and material identities before selecting a candidate or requesting
another segmentation. The selected contours enter the alignment and
device-construction stages.

\subsection{Registration across assembly stages}\label{sec:supp-registration}

\qlaybot{} registers the source microscope images to the assembled
heterostructure and maps the assembled view to the chip layout,
preserving material identity across transfer and placing selected
boundaries in lithographic coordinates.

For images sharing a substrate, the alignment program matches local
features using SIFT and estimates a similarity transform from RANSAC
inliers. For stamp-to-stack registration, the agent selects an
approximate orientation from overlay images before the program optimizes
rotation, scale, and translation against a contour-distance objective
with penalties for material extending outside the observed footprint or
image. Differential evolution initializes bounded L-BFGS-B refinement,
and the supplied transfer convention determines reflection of the stamp
view.

The second alignment detects chip markers in the assembled-stack image
and matches them with marker geometry in the GDS template. The program
tests similarity transforms, refines accepted matches, and converts the
aligned material contours to micrometers. For a source point \(p\), the
resulting chip position follows the composed mappings
\[
p_{\mathrm{chip}}=T_{\mathrm{stack}\rightarrow\mathrm{chip}}\!\left(T_{\mathrm{source}\rightarrow\mathrm{stack}}(p)\right).
\]
The program writes the final material contours directly into the chip
coordinate system. The same contours guide channel, contact, gate, and
lead placement.

\subsection{Device construction and routing}\label{sec:supp-routing}

The agent uses the overlap between graphene and the backgate to choose
the channel location and orientation. The exclusive conductive regions
determine where the corresponding contact patches can lie. Geometry
tools construct the mesa, arms, topgate, contact patches, and other
device features as GDS polygons.

The chip template provides fixed bonding pads. The routing tools first
assign inner contact points to outer pads. They sort both groups by
angle and seek an order-preserving assignment with minimum total
distance. For contact \(i\) and candidate pad \(j\), the dynamic program
uses
\[
D_{i,j}=d(i,j)+\min_{k<j}D_{i-1,k},
\]
where \(d(i,j)\) is the distance between the two connection points. The
agent can also specify a contact--pad assignment explicitly.

The router converts obstacles and clearance requirements into a weighted
grid and finds minimum-cost lead paths. Completed leads constrain the
placement of later leads. The agent can modify assignments, widths,
obstacles, and routing regions, or provide explicit path points. The
tools insert the resulting lead geometry into the layout.

\subsection{Inspection and revision}\label{sec:supp-inspection}

KLayoutClaw exposes geometric operations and checks to the agent through
the Model Context Protocol. Checks report channel placement,
contact-material overlap, clearance, routing connectivity, and lead
intersections. The agent uses their measurements and locations to revise
the selected contours, device features, or routes. The design is saved
as GDSII for external evaluation against the reference specimen.

The tool implementation and associated design instructions are available
in \href{https://github.com/caidish/KlayoutClaw}{KLayoutClaw}.

\section{Representative device designs}\label{sec:supp-designs}

\subsection{QH06: local device placement with an oversized graphene contour}\label{sec:supp-qh06}

In the QH06 end-to-end design, an agent using Claude Opus 5 inferred a
graphene contour covering 99.2\% of the reference graphene. Its area was
5.3 times the reference area. The contour had an intersection-over-union
of 0.1873. The graphite contour reached an overlap value of 0.8751.

\begin{figure}[!htb]
\centering
\includegraphics[width=\linewidth]{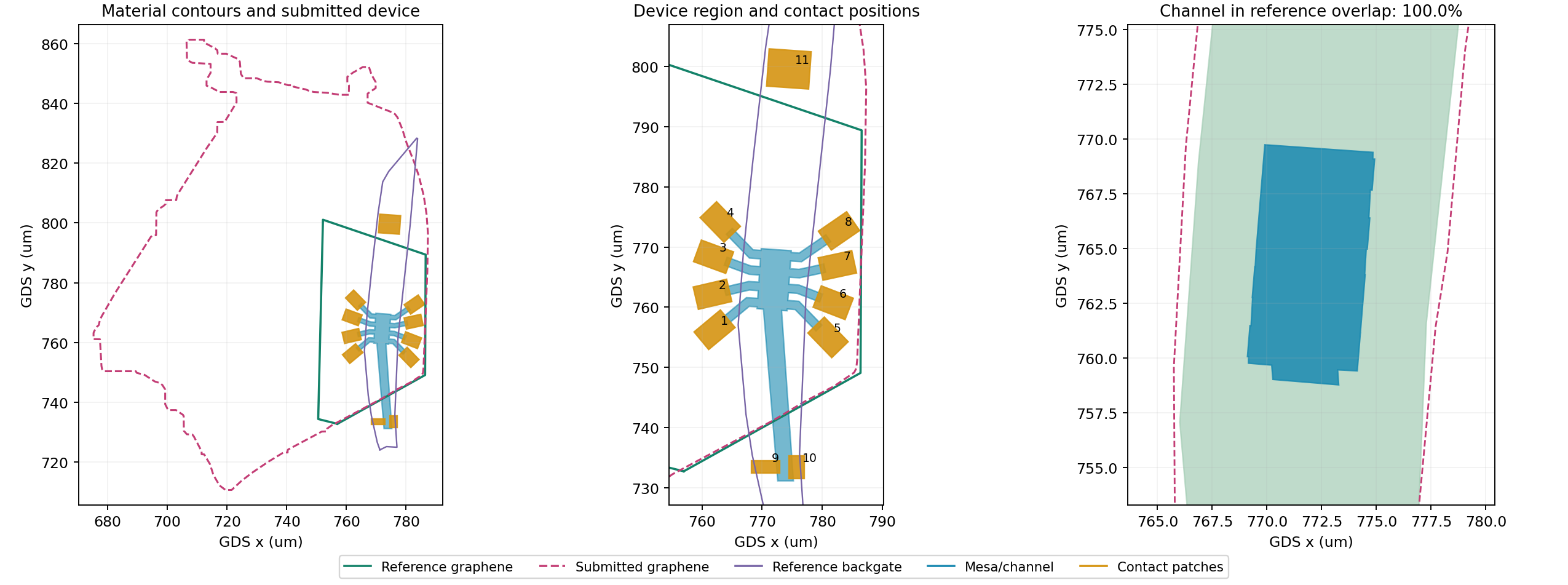}
\caption{QH06 device and reference geometry. Left: submitted and
reference material contours. Center: contact positions. Right: central
channel in the reference conductive-layer overlap.}
\label{fig:supp-qh06}
\end{figure}

The agent found insufficient contact space on one side of its proposed
channel. It measured the available graphene area, shifted the channel,
and shortened the connecting arms. In the submitted GDS layout
(Fig.~\ref{fig:supp-qh06}), the central channel lies inside the
reference graphene--backgate overlap, and all eight ohmic contact
centers occupy the reference graphene-only region. The channel,
contact-material, and connectivity checks each received full credit.
The device score was 0.9625. The flake-detection metric applied to the
same submitted contours gave 0.6667.

\subsection{ml14: contact verification against an incorrect material contour}\label{sec:supp-ml14}

In the ml14 end-to-end design, an agent using Claude Sonnet 5
interpreted an elongated feature in the pre-transfer stamp image as
graphene and top hBN. It constructed a mesa with four ohmic contacts at
each end and connected the contacts through arms.

\begin{figure}[!htb]
\centering
\includegraphics[width=0.85\linewidth]{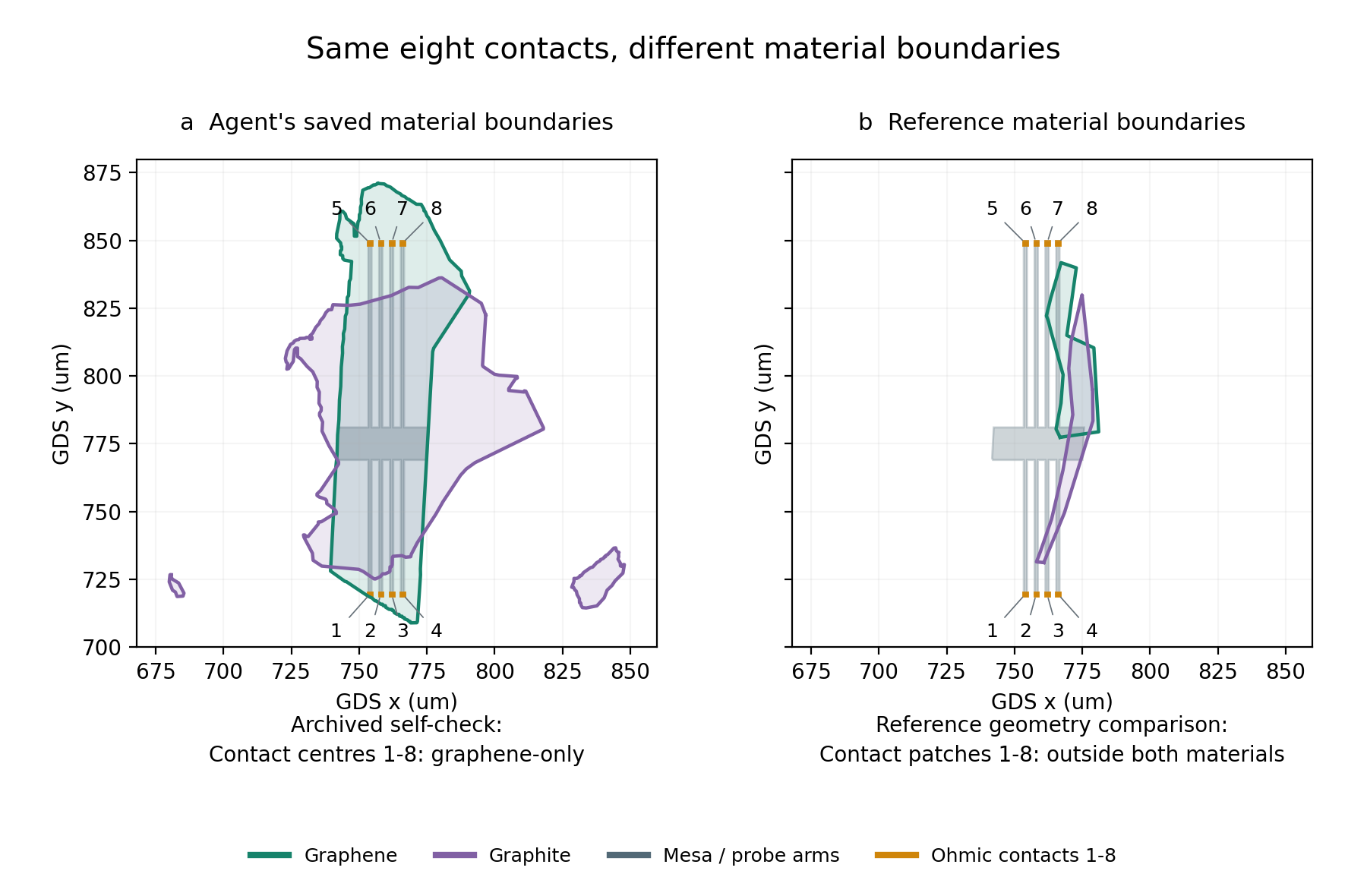}
\caption{ml14 contact checks and reference geometry. Left: the agent's
stored material boundaries and contact-check results. Right: expert
reference contours and the eight submitted contact patches.}
\label{fig:supp-ml14}
\end{figure}

The agent's checking script reopened the saved GDS layout and tested the
eight contact centers against its stored graphene and backgate contours.
All eight contacts passed that local material test. The agent also
identified and repaired a gate-connection crossing. Evaluation against
the expert contours placed all eight contact patches outside both
reference conductive materials. The device received a score of 0.250135.
The contact-check result and the reference overlay appear in
Fig.~\ref{fig:supp-ml14}.

\end{document}